\documentclass[prd,twocolumn,aps,showpacs,nofootinbib,nobibnotes,superscriptaddress,preprintnumbers]{revtex4}
\usepackage{epsfig}
\usepackage{graphics}
\usepackage{bm}
\usepackage{color}
\usepackage{dcolumn}   
\usepackage[spanish,english]{babel}
\usepackage{bm}     
\usepackage{bbm}       
\usepackage{amssymb}  
\usepackage{amsmath}
\usepackage{latexsym}
\usepackage{float}
\usepackage{ifthen}
\usepackage{caption,subfig}
\usepackage{enumerate}
\usepackage{url}
\usepackage{caption,subfig}
\usepackage{amsopn}
\usepackage{hyperref}

\usepackage{amsfonts}
\usepackage{multirow}
\usepackage{array}
\usepackage{rotating}

\usepackage{ulem}
\def\clap#1{\hbox to 0pt{\hss#1\hss}}

\def\bea{\begin{eqnarray}}
\def\eea{\end{eqnarray}}
\def\be{\begin{equation}}
\def\ee{\end{equation}}

\renewcommand{\geq}{\geqslant}

\begin{document}

\title{On the Scalar-Vector-Tensor Gravity: Black Hole, Thermodynamics and Geometrothermodynamics}

\author{Phongpichit Channuie}
\email{channuie@gmail.com}
\affiliation{ School of Science, Walailak University, Thasala, \\Nakhon Si Thammarat, 80160, Thailand}

\author{Davood Momeni} \email{davood@squ.edu.om}
 \affiliation{{Department of Physics, College of Science, Sultan Qaboos University,\\P.O. Box 36, P.C. 123,  Muscat, Sultanate of Oman}}

\date{\today}

\begin{abstract}

Recently, a new class of modified gravity theories formulated via an additional scalar and vector field on top of the standard tensor field has been proposed. The direct implications of these theories are expected to be relevant for cosmology and astrophysics. In the present work, we revisit the modified framework of the scalar-vector-tensor theories of gravity. Surprisingly, we discover novel metric function for the black hole solutions. We also investigate the semi-classical thermodynamics of the black holes and study the thermodynamic properties of the obtained solutions. Moreover, we quantify the entropy and the temperature of the new black hole and also calculate the heat capacity. Finally, we also apply the formalism of the geometrothermodynamics to examine thermodynamic properties of the new black hole. This formalism yields results consistent with those obtained from the usual thermodynamic implementation.   

\end{abstract}


\maketitle

\section{Introduction}
It was strongly evident that the predictions from the Einstein
theory are in excellent agreement with the gravitational observations. However, there exist some physical indications implying that the standard General Relativity could be in principle modified. The simplest extension involves a scalar field. In this context, however, the theories of gravity may
result a higher-order time derivative in the equations of motion, and consequently these theories may encounter Ostrogradski instabilities. 

In some cases, the higher-order time derivatives in the equations of motion can be  reduced to a second-order one \cite{Zumalacarregui:2013pma}. As is well known, the most general second order, covariant scalar-tensor theories leads to Horndeski interactions \cite{Horndeski:1974wa}. This class of theories have proved to be instability-free which gives rise to equations of motion no higher than second order in time derivative.

Nevertheless, the modification in the scalar sector is not the only option. The inclusion of an additional vector field into the gravity sector is another viable possibility. Here we can come up with the most general vector-tensor theories with second order equations of motion. As discussed in Ref.\cite{Horndeski:1976gi}, only one additional coupling of the vector field to the double dual Riemann tensor is possible \cite{Horndeski:1976gi} by imposing gauge invariance on the vector field. Here one also assumes that when the curvature tensor vanishes, we recover Maxwell's equations as the equations governing the vector fields. It is worth noting that when abandoning gauge invariance we recover the general vector-tensor theories known as the generalised Proca theories \cite{Heisenberg:2014rta}. 

It was noticed in Ref.\cite{Jimenez:2016isa} that there are two new genuine purely intrinsic vector interactions without scalar counterpart in the theories. One of these interactions non-minimally couples to the double dual Riemann tensor (see also \cite{VectorTensorTheories}). It was claimed in Ref.\cite{Heisenberg:2018acv} that these vector-tensor gravity theories may yield very rich cosmological \cite{VTcosmology} and astrophysical \cite{VTastrophys} implications. Impressively, it seems like the unification of the Horndeski theories and the generalized Proca theories \cite{deFelice:2017paw} can be achieved via these scalar-vector-tensor (SVT) gravity theories (see also Refs.\cite{Heisenberg:2018} for new hairy black holes solutions). Very recently, the application to dark energy \cite{Kase:2018nwt} has been implemented.

As pointed out in Ref.\cite{Heisenberg:2018acv}, these theories may have rich applications to cosmology, especially to the early universe and dark matter phenomenology. Moreover, the SVT theories were constructed for both $U(1)$ gauge-invariant and broken gauge-invariant cases. In this work, we will focus on the gauge-invariant case. There exists another different class of the modified gravity theory, called MOG, which can alternatively explain the flat rotation curve of galaxies without invoking cold dark matter particles \cite{Moffat:2005si} (see also recent examination \cite{ra}).

This paper is organized as follows. In Sec.\ref{s2}, we briefly review the modified framework of the scalar-vector-tensor theories of gravity. In Sec.\ref{s3}, we examine the BH solutions and quantify the metric function for the new black hole. In Sec.\ref{s4}, we also follow an approach of the semi-classical thermodynamics of the black holes and study the relevant thermodynamic properties of the obtained solutions. In Sec.\ref{s5}, we alternatively apply the formalism of the geometrothermodynamics to examine thermodynamic properties of the new black hole. The last section is devoted to remarks and conclusions.

\section{Gauge-Invariant Scalar-Vector-Tensor Theories revisited}
\label{s2}
We in this section revisit the modified gravity theories recently proposed by Ref.\cite{Heisenberg:2018acv} called scalar-vector-tensor gravity theories (SVT). Here the unification of the Horndeski theories and the generalized Proca ones can be achieved. In the gauge invariant platform, we follow the model proposed by Ref.\cite{Heisenberg:2018acv}. Here the Lagrangian densities of the genuine scalar-vector-tensor interactions take the following form: 
\begin{eqnarray}\label{genLagrangianSVTnoGauge}
\mathcal{L}^{2}_{\rm SVT}&=&f_2(\pi,X,F,\tilde{F},Y), \\
\mathcal{L}^{3}_{\rm SVT}&=& {\cal M}_{3}^{\mu\nu}\nabla_\mu\nabla_\nu\pi,\label{eq2}\\
\mathcal L^{4}_{\rm SVT} & = & {\cal M}_{4}^{\mu\nu\alpha\beta}\nabla_\mu\nabla_\alpha\pi\nabla_\nu\nabla_\beta\pi,\nonumber\\
&&+f_{4}(\pi,X)L^{\mu\nu\alpha\beta}F_{\mu\nu}F_{\alpha\beta},\label{eq3}
\end{eqnarray}
where the functions $X=-\frac12(\partial\pi)^2$, $F=-F_{\mu\nu}F^{\mu\nu}/4$, ${\tilde F}=-F_{\mu\nu}{\tilde F}^{\mu\nu}/4$ and $Y=\nabla_\mu\pi\nabla_\nu\pi F^{\mu\alpha}F^\nu{}_\alpha$ with the gauge-invariant field strength and its dual given by $F_{\mu\nu}=\nabla_{\mu}A_{\nu}-\nabla_{\nu}A_{\mu}$ and ${\tilde F}_{\mu\nu}=\epsilon^{\mu\nu\alpha\beta}F_{\alpha\beta}/2$, respectively. Here $\epsilon^{\mu\nu\alpha\beta}$ is the anti-symmetric Levi-Civita tensor. The rank-2 tensor
${\cal M}^{\mu\nu}_{3}$ in Eq.(\ref{eq2}) is of the form
\begin{eqnarray}
{\cal M}^{\mu\nu}_{3}= \Big(f_{3}(\pi,X)g_{\rho\sigma}+{\bar f}_{3}(\pi,X)\nabla_{\rho}\pi\nabla_{\sigma}\pi\Big)\tilde{F}^{\mu\rho}\tilde{F}^{\nu\sigma}\,,
\end{eqnarray}
where $f_{3}$ and ${\bar f}_{3}$ are functions of $\pi$ and $X$. Note that the rank-4 tensor  $\mathcal{M}_{4}^{\mu\nu\alpha\beta}$ in Eq.(\ref{eq3}) is given by
\begin{equation}
\mathcal{M}_{4}^{\mu\nu\alpha\beta}=\Big( \frac12f_{4,X}(\pi,X)+{\tilde f}_{4}(\pi)\Big)\tilde{F}^{\mu\nu}\tilde{F}^{\alpha\beta}\,,
\end{equation}
where $L^{\mu\nu\alpha\beta}$ is the double dual Riemann tensor formulated with the help of the Riemann tensor $R_{\rho\sigma\gamma\delta}$ as
\begin{eqnarray}
L^{\mu\nu\alpha\beta}=\frac14 \epsilon^{\mu\nu\rho\sigma}\epsilon^{\alpha\beta\gamma\delta} R_{\rho\sigma\gamma\delta}\,,
\end{eqnarray}
Moreover, we can then formulate theories beyond scalar-vector-tensor by performing disformal transformations. This allows us to generalize the interactions in ${\cal L}^{2,3,4}_{\rm SVT}$. Note that there were some previous works on the implications of disformal transformation on cosmology \cite{disformal}. However, this generalization may be presumed for the future work. In the following sections we will figure out how the present theories provide some particularly physical applications to cosmology.   

\section{Black hole solutions}
\label{s3}
\begin{figure}
	\includegraphics[width=8.0cm]{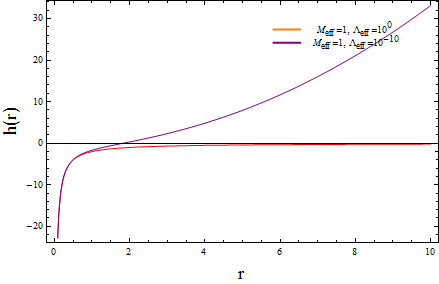}
    \includegraphics[width=8.0cm]{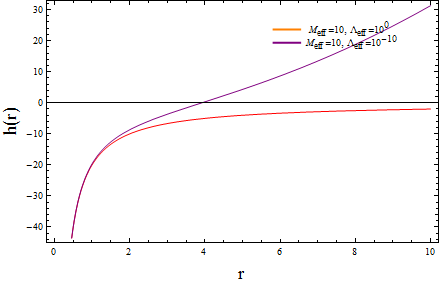}
	\caption{We plot $h(r)$ versus $r$ for $M_{\rm eff}=1,\,\Lambda_{\rm eff}=1$ and  $M_{\rm eff}=1,\,\Lambda_{\rm eff}=10^{-10}$  displayed on the upper-panel and for $M_{\rm eff}=10,\,\Lambda_{\rm eff}=1$ and  $M_{\rm eff}=10,\,\Lambda_{\rm eff}=10^{-10}$ on the lower-panel.} \label{frmbe}
\end{figure}

As suggested in Ref.\cite{Heisenberg:2018acv}, new black hole (and perhaps neutron star) solutions may be accommodated as one of the crucial implications of these SVT theories. It is reasonable to begin our investigation by firstly considering the black hole solutions of the present theories. In order to examine these possible solutions, let us consider the following Lagrangian with metric signature $(-,+,+,+)$:
\begin{equation}\label{genLagrangian}
\mathcal{L}=\sqrt{-g}\left(\mathcal{L}_{\rm gravity}+\mathcal{L}^4_{\rm SVT}\right)\,,
\end{equation}
where $\mathcal{L}_{\rm gravity}=\frac{M_{\rm Pl}^2}{2}R+G_2(\pi,X)$. Here, we consider the case where the solutions are static and spherically symmetric with the metric background:
\begin{equation}
ds^2=-f(r)dt^2+h^{-1}(r)dr^2+r^2d\Omega^2,\label{matrix}
\end{equation}
where the scaleron ($\pi$), gauge field ($A_{\mu}$) and metric function ($f$) are all functions of $r$, that is to say $\pi=\pi(r),A_\mu=(A_0(r),0,0,0)$ and $f=f(r)$. In terms of these field configurations, the Lagrangian $\mathcal{L}^4_{\rm SVT}$ takes the form \cite{Heisenberg:2018acv}:
\begin{equation}
\mathcal{L}^4_{\rm SVT}=\frac{h A_0'^2(-4f_4(h-1)+h^2\pi'^2(2\tilde{f}_4+f_{4,X}))}{r^2f}, \label{fullL}
\end{equation}
where primes denote derivative with respect to $r$. It is straightforward to derive  equations of motion for $\pi,\,A_{0},\,h$ and $f$ using standard techniques. However, it was expected from Ref.\cite{Heisenberg:2018acv} that different types of new hairy black hole solutions will be obtained in these theories. This is so since the equations of motion contributing from $\mathcal{L}^4_{\rm SVT}$ are different from those obtained from the generalized Proca theories. Therefore, it is reasonable to figure out weather the new hairy black hole solutions can be quantified regarding these SVT theories.  
In order to examine the solutions for the full Lagrangian $\mathcal{L}_{\rm BH}=\mathcal{L}^4_{\rm gravity}+\mathcal{L}^4_{\rm SVT}$, we first consider those obtained from the $A_0$ sector. We obtain in this case 
\begin{equation}
\frac{h A_0'(-4f_4(h-1)+h^2\pi'^2(2\tilde{f}_4+f_{4,X}))}{r^2f}=p_a\,,
\end{equation}
where $p_a=\frac{\partial \mathcal{L}^4_{\rm SVT}}{\partial A'_0}$ denotes conserved conjugate momentum of the system associated with the field configuration. In metric background $X=\frac{1}{2f}\pi'^2$. What we are interested in is the shift symmetric black holes where $f_{4}=0$ and we choose 
$\tilde{f}_{4}=X$. Regarding these conditions, the field configuration reduces to
\begin{equation}
\frac{4h^3 A_0' X}{r^2}=p_a\,. \label{reduce}
\end{equation}
It is naively suggested from the above equation that one can obtain the solutions by considering some particular choices of configuration. In this present analysis, we opt $f=h$ . Considering this shift symmetric model, a field profile $A_{0}$ can be simply derived:
\begin{equation}
A_0=\frac{p_a}{4}\int_{\bar r}\frac{\bar r^2}{h(\bar r)^3 X(\bar r)}d\bar r\label{field}
\end{equation}
where we need the scalar field profile $\pi (r)$. Let us next figure out what the metric function $h$ looks like. To this end, we take a variation of the total action 
(\ref{genLagrangian}) w.r.t $f$ or $h$. With the help of Eq.(\ref{reduce}), we come up with the following linear first-order differential equation for $h$:
\begin{eqnarray}
-b r^2 + M_{\rm Pl}^2 (h + r h')=0
\label{hnon}
\end{eqnarray}
where the argument of $h$ is understood and we have assumed that $G_2(\pi,X)=aX+b$ preserving the shift symmetry. Regarding the above differential equation, here we figure out the solutions after performing an integration:
\begin{eqnarray}
h=\frac{\Lambda _{\rm eff} r^2}{3}-\frac{2M_{\rm eff}}{r},\label{hr}
\end{eqnarray}
where $b,M_{\rm eff}$ are integration constants and we have defined a (cosmological) constant as:
\begin{eqnarray}
&& \Lambda_{\rm eff}=\frac{b}{ M_{\rm Pl}^2},\ \ b>0.
\end{eqnarray}
The plots of metric function $h(r)$ can be seen in Fig.\ref{frmbe} with some particular conditions. It is worth noting that the above metric function coincides with that of the planar de Sitter-Schwarzschild which corresponds to the gravitational field of a  non-rotating, spherically symmetric body of mass $M_{\rm eff}$ in de Sitter spacetime.
Using (\ref{genLagrangian}), now we turn to quantify the scaleron profile $\pi$ and it is obtained by examining possible solutions for the equation of motion such that $p_{\pi}=\frac{\partial \mathcal{L}_{\rm SVT}}{\partial \pi'}=\mbox{constant}$ (because $\pi$ is a cyclic variable like a field $A_0$).
\begin{eqnarray}
&& a + 2 c X + \frac{4 h^3 A_0'^2}{r^2}= \frac{p_{\pi}}{r^2},\label{pi}
\end{eqnarray}
Using (\ref{reduce}) we can eliminate $X$ and obtain
\begin{eqnarray}
a+\frac{p_a^2}{4r^2u^2h}+\frac{cu}{h}=\frac{p_{\pi}}{r^2},\label{pi1}
\end{eqnarray}
with $u=\pi'^2$. This unique real solution for $u\in\mathcal{R
}^{+}$  for u and exact solution can be obtained via
\begin{widetext}
\begin{eqnarray}
\pi'^2  =&&\frac{\sqrt[3]{\Sigma_{1}+\sqrt{\left(\Sigma_{1}+ \Sigma_{2}\right)^2-16384 h^6
			\left(p_{\pi}-a r^2\right)^6}+ \Sigma_{2}}}{12
	\sqrt[3]{2} c r^2}\nonumber\\&&+\frac{4 \sqrt[3]{2} h^2 \left(p_{\pi}-a r^2\right)^2}{3 c r^2
	\sqrt[3]{\Sigma_{1}+\sqrt{\left(\Sigma_{1}+ \Sigma_{2}\right)^2-16384 h^6 \left(p_{\pi}-a
			r^2\right)^6}+ \Sigma_{2}}}+\frac{h
	\left(p_{\pi}-a r^2\right)}{3 c r^2}
\label{pis}
\end{eqnarray}
\end{widetext}
where
\begin{eqnarray}
&&\Sigma_{1}\equiv -128 a^3 h^3 r^6+384 a^2 h^3 p_{\pi} r^4,\nonumber\\&&\Sigma_{2}\equiv -384 a h^3 p_{\pi}^2 r^2-432 c^2 p_a^2 r^4+128 h^3 p_{\pi}^3.\nonumber
\label{pis}
\end{eqnarray}
In case of the asymptotic limit, we obtain for $\pi(r)$ as follows:
\begin{eqnarray}
&&\pi(r)=\int_{ r}\pi'(\bar r)d\bar r.
\end{eqnarray}
We obtained the BH solutions with $f=h$ and $A_{0}\neq 0$ and $\pi\neq 0$ as suggested in \cite{Heisenberg:2018acv}. However our framework, based on the SVT theories, constitutes the natural template for other constraints when one want to quantify the BH solutions.

\section{Thermodynamics}
\label{s4}
\begin{figure}
	\includegraphics[width=8.0cm]{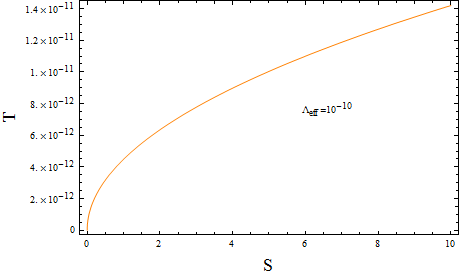}
	\includegraphics[width=8.0cm]{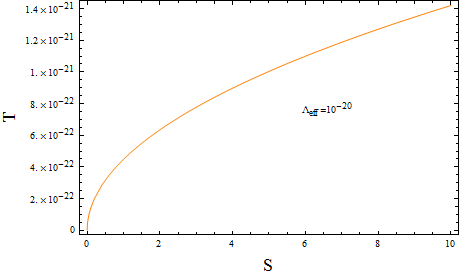}
	\caption{We plot the temperature $T$ as a function of the entropy $S$, with $\Lambda_{\rm eff}=10^{-10}$ displayed on the upper-panel and for $\Lambda_{\rm eff}=10^{-20}$ on the lower-panel..} \label{BHtem}
\end{figure}

In our analysis, we will follow the semi-classical approach to the thermodynamics of black holes, initiated by Hawking \cite{hawking1} and developed subsequently by other authors \cite{davies2}. In this section, we turn to study the thermodynamic properties of the solutions. To begin with, we will express the effective mass $M_{\rm eff}$ in terms of the radius of the events (outermost) horizon $r_+$ and the charge $Q_{\rm eff}$. This can be achieved by equating $g_{00}=f(r)$ to zero. Here we obtain
\begin{equation}
M_{\rm eff}=\frac{\Lambda_{\rm eff}r_{+}^3}{6}.\label{mass1}
\end{equation}
It is worth noting that there have been various equivalent approaches of obtaining the Hawking temperature (see some of them \cite{ford,hawking2,davies,davies2,gerard3,kanti,k21,r5,robinson,jacobson}). In the following, we compute the Hawking temperature of the black hole on the event horizon by using the definition of surface gravity \cite{ford,hawking2,gerard3,kanti,k21,r5,robinson,jacobson}, $\kappa$, such that
\begin{equation}
\kappa =\left[\frac{|g_{00}^{\prime}|}{2\sqrt{-g_{00}g_{11}}}\right]_{r=r_{+}}\, ,\label{10}
\end{equation}
which is evaluated at the radius of the events horizon. Therefore, the Hawking temperature can be usually related to the surface gravity via the relation for the case of the metric function obtained in Eq.(\ref{hr}):
\begin{equation}
T=\frac{\kappa}{2\pi}=\frac{f'(r_{+})}{4\pi}= \frac{r_{+}\Lambda_{\rm eff}}{4\pi}.\label{11}
\end{equation}
Using the line element given in Eq.\ref{matrix}, it is simple to show that $A=4\pi r_{+}^{2}$. In the context of standard gravity, the so-called area law can be still used to examine the entropy of the black holes. Therefore, the entropy of the black hole takes the following explicit form \cite{bardeen}
\begin{equation}
S=\frac{1}{4}A=\pi r_{+}^{2}.\label{s1} 
\end{equation}
In terms of the entropy, the temperature takes the form:
\begin{equation}
T= \frac{S^{1/2}\Lambda_{\rm eff}}{4\pi^{3/2}}.\label{ts}
\end{equation}
It is simple to verify that the temperature $T$ will be always positive. The temperature increases as a function of entropy $S$. Then, as the entropy increases, the temperature becomes an increasing function as a power-law. The behavior of the temperature as a function of the entropy is displayed in Fig.\ref{BHtem}.
What we are going to do next is to verify the validity of the first law of BH thermodynamics for our obtained solutions. Here we consider the first law of thermodynamics in differential form
\begin{equation}
dM_{\rm eff}=TdS+P d\Lambda_{\rm eff},\label{plt}
\end{equation}
where $dM_{\rm eff}$ and $dS$ are 
\begin{eqnarray}
dM_{\rm eff} &=&\frac{\Lambda_{\rm eff}r_{+}^2}{2}dr_{+}+\frac{r_{+}^3}{6}d\Lambda_{\rm eff},\\
dS &=& 2\pi r_{+}dr_{+}.
\end{eqnarray}
Note that the above equation may be extended by adding other terms, such as $VdP$ \cite{Hendi:2018hdo}. We can properly establish the study of the thermodynamics of the BH system. From Eq.(\ref{s1}), we get  $r_+=\sqrt{S/\pi}$ and insert this into (\ref{mass1}) to obtain the BH mass as the function of extensive parameters $S$ and $\Lambda_{\rm eff}$:
\begin{eqnarray}
M_{\rm eff}(S,\Lambda_{\rm eff})=\frac{\Lambda_{\rm eff}S^{3/2}}{6\pi^{3/2}}.\label{mass2}
\end{eqnarray}
It is matter of calculating to show that
\begin{eqnarray}
\left(\frac{\partial M_{\rm eff}(S,\Lambda_{\rm eff})}{\partial S}\right)_{\Lambda_{\rm eff}}=T\,,\,\,\left(\frac{\partial M(S,\Lambda_{\rm eff})}{\partial \Lambda_{\rm eff}}\right)_{S}=P.\label{mass21}
\end{eqnarray}
From (\ref{plt}), we obtain the following expressions: 
\begin{eqnarray}
&&\left(\frac{\partial M_{\rm eff}}{\partial S}\right)_{\Lambda_{\rm eff}}=
\frac{\Lambda_{\rm eff}S^{1/2}}{4\pi^{3/2}}
,\label{eqe01}\\
&&\left(\frac{\partial M_{\rm eff}}{\partial \Lambda_{\rm eff}}\right)_{S}=\frac{S^{3/2}}{6\pi^{3/2}}
\label{eqe}.
\end{eqnarray}
Exactly, in terms of $r_{+}$ our result in Eq.(\ref{eqe01}) is matched with that of (\ref{11}). 
Before ending this section, it is reasonable to check thermal stability of the obtained BH solutions. In this regard, as mentioned in Refs.\cite{Hendi:2018hdo,Dehghani:2018qvn} , in order to examine the stability conditions we should consider the sign of heat capacity (positive or negative). The heat capacity at constant effective cosmological constant is characterized by the following relation:  
\begin{eqnarray}
C_{\Lambda_{\rm eff}}=\frac{T}{\Big(\frac{\partial^{2} M}{\partial S^{2}}\Big)_{\Lambda_{\rm eff}}},
\end{eqnarray}
where $T$ has already been derived in Eq.(\ref{11}). It was proved in Ref.\cite{Dehghani:2018qvn} that with $T>0$  the positivity of heat capacity are sufficient to ensure the local stability of the black hole. Consider (\ref{s1}) and (\ref{mass2}), we can simply show that the denominator of the heat capacity reads
\begin{eqnarray}
\Big(\frac{\partial^{2} M}{\partial S^{2}}\Big)_{\Lambda_{\rm eff}}=\frac{\Lambda_{\rm eff}}{8\pi^{3/2}S^{1/2}}.
\end{eqnarray}
Note here that the thermal stability conditions are based on the sign of heat capacity. As a result, what we are looking for is the root and divergence points of the heat capacity. For the divergence points of the BH heat capacity, we consider the real roots for $(\partial^{2}M/\partial S^{2})_{\Lambda_{\rm eff}}=0$. In terms of the entropy $S$, the heat capacity takes the form:  
\begin{eqnarray}
C_{\Lambda_{\rm eff}}=2S. \label{BHcap}
\end{eqnarray}
It was mentioned in Ref.\cite{da} that the second order phase transitions
will take place at the points where the heat capacity diverges. In this case, that yields
\begin{eqnarray}
S=\infty. \label{BHcap1}
\end{eqnarray}
This may mimic that a black hole with infinite entropy means no further change can take place at this point. More detailed discussion regarding the thermal stability can be found in Refs.\cite{Hendi:2018hdo,Dehghani:2018qvn} and references therein. 

\section{Geometrotermodynamic}
\label{s5}
So far, it was shown that the geometrical approach to thermodynamics can be achievable and useful. In this present work, the contact geometry approach will be adopted to study a physical thermodynamic system. This method is commonly known as geometrothermodynamics (GTD). The formalism of the GTD allows us to construct a mathematical space where all the thermodynamic quantities can be physically defined. 

In order to examine the thermodynamics of the system, we start by introducing the $(2n+1)-$dimensional  phases space $\cal{T}$, parametrized by coordinates $Z^{C}=\{\Phi,\,E^{a},\,I^{a}\}$ where $C=0,1,...,2n$ and $a=1,...,n$. Here $n$ represents the number of thermodynamics degree of freedom of the system. Suppose that the $\cal{T}$ is differentiable manifold and possesses a non-degenerated metric $G_{AB}(Z^{C})$. Assume that the Gibbs 1-form $\Theta=d\Phi -\delta_{ab}I^{a}dE^{b}$, where $\delta_{ab}$ is the Kronecker delta, satisfies the condition  $\Theta\wedge \left(d\Theta\right)^{n}\neq 0$. Here we call the set $\{{\cal T} , \Theta , G\}$ a contact Riemann manifold \cite{hermann}.

To successfully describe the system, we also define the equilibrium space which is the subspace of $\cal {T}$ defined via the smooth map $\varphi:{\cal E}\rightarrow{\cal T}$, with $\Phi\equiv \Phi (E^{a})$, under the condition 
\begin{eqnarray}\label{cond1}
\varphi^{*}(\Theta)\equiv 0 \Rightarrow \left\{\begin{array}{ll}
d \Phi=\delta_{ab}I^{a}dE^{b}\,\,,\\ 
\frac{\partial\Phi}{\partial E^{a}}=\delta_{ab}I^{b}\,\,, \end{array}\right.
\end{eqnarray}
where $\varphi^{*}$ is the pullback of $\varphi$ and the $n$-dimensional space $\cal {E}$ spans by the coordinates $E^{a}$. Here the first expression of (\ref{cond1}) yields the first law of thermodynamics, and the second one possesses the condition for thermodynamic equilibrium. A required condition of the second law of thermodynamics is followed by
\begin{equation}
\pm\frac{\partial^{2}\Phi}{\partial E^{a}\partial E^{b}}\geq 0\; ,\label{segunda}
\end{equation}
where the chosen thermodynamic potential implies the sign of (\ref{segunda}), i.e. $+$ or $-$. In the mass case, we get the positive sign ($+$) and for the entropy case we instead get the negative one ($-$). Note that ${\cal E}$ is a submanifold of points where the first law and the equilibrium conditions hold. We can also identify the set $\{E^{a}\},\,\Phi=\Phi(E^{a})$, and the coordinates $\{I^{a}\}$ as the extensive thermodynamic variables, the thermodynamic potential, and the intensive quantities, respectively.

What we require next is to introduce the metric $g$ in ${\cal E}$ whose properties are independent of the choice of thermodynamic potential. Regarding to the requirement, we can construct such a metric via $g=\varphi^{*}(G)$ with $G$ being a metric in ${\cal T}$ preserving the Legendre invariance. In general, there may exist metrics $G$ that are not Legendre invariance. However, the most general metric invariant under total Legendre transformations can be expressed in the following form:  
\begin{eqnarray}
G &=&G_{AB}dZ^{A}dZ^{B}\nonumber\\&=&\Theta^{2}+\left(\delta_{ab}E^{a}I^{b}\right)\left(\eta_{ab}dE^{a}dI^{b}\right)\; \label{mt}\;,
\end{eqnarray}
where $\eta_{ab} =\{\pm 1,1,...,1\}$ for which the second order phase transition implies $\eta_{ab} =\{ -1,1,...,1\}$. In order to constrcut the metric in $\cal E$, we adopt $g=\varphi^{*}(G)$ such that:
\begin{eqnarray}
g &=&\varphi^{*}(G) = g_{ab}dE^{a}dE^{b} \nonumber\\&& =\frac{\partial Z^{A}}{\partial E^{a}}\frac{\partial Z^{B}}{\partial E^{b}}G_{AB}dE^{a}dE^{b},\nonumber\\&
 & =\left( E^{c}\frac{\partial\Phi}{\partial E^{c}}\right) \left( \eta_{ad}\delta^{di}\frac{\partial^{2}\Phi}{\partial E^{i}E^{b}}\right)dE^{a}dE^{b}\; . \label{me}
\end{eqnarray}
What we require is to have the contact structure ${\cal T}$ invariant under the contact transformation which consequently implies that the geometrical properties of $G$ are independent of the choice of the thermodynamic potential. Having required the metric to be Legendre invariant, we write \cite{quevedo}
\begin{align}
\Phi=\widehat{\Phi}-\delta_{ab}\widehat{E}^{a}\widehat{I}^{b}\; , \; E^{a}=-\widehat{I}^{a}\; , \; I^{b}=\widehat{E}^{b}\; .\label{tl}  
\end{align}
Below we will study the phase transitions and stability of the thermodynamic system. In so doing, we need first to determine the curvature scalar $R$ using the metric (\ref{me}). , which provides two implications: (I) Whether there exist thermodynamic interaction and phase transition, and (II) at which point of the  space, thermodynamic equilibrium states can occur.

For a given set of variables $\{X^{1},X^{2},...,X^{n}\}$ {\it viz.}, charges, temperature, entropy, etc., the local stability of the underlying state-space configuration requires the heat capacity to be positive \cite{bellucci}:
\begin{eqnarray}
\{g_{ii}(X^{i})>0\,;\,\,\forall i=1,2,...,n\}\,. 
\end{eqnarray}
As a result, the following equations must be simultaneously satisfied maintaining the stability condition: $p_{0}>1,\,\,p_{1}>0,\,\,...,\,\,p_{n}=\det\left[ g_{ab}\right]>0$ (see \cite{bellucci} for more details). 

Bear in mind that the model we are considering is non-rotating black holes. It is worth noting that another mechanism for examining the global stability (instability) is the Helmholtz free energy. However, we shall not include it in our present analysis. In our present case, we can also define the Gibbs potential as 
\begin{align}
G(T,\Phi)=M(S,\Lambda_{\rm eff})-TS-P \Lambda_{\rm eff}\;,\label{gibbs}
\end{align}
and the global stability ensures the condition
\begin{align}
G(E^{a})<0\; ,\; \forall E^{a}\; (E^{a}\in I(E^{a}))\; . \label{eggibbs} 
\end{align}
In order to quantify the global stability of the thermodynamic system, we consider the thermodynamic variables of the system which basically include the mass $M$ and the cosmological constant  $\Lambda_{\rm eff}$. Note that other variables including the entropy $S$, the temperature $T$ and the electric potential scalar $A_{0}$ can be basically defined from these two parameters.   

Here we use the BH mass representation $M(S,\Lambda_{\rm eff})$ derived in Eq.(\ref{mass2}) to study the thermodynamic description of the system. In the present analysis, the entropy $S$ and the charge $\Lambda_{\rm eff}$ are the extensive variables ($E^{a}$); whilst the temperature $T$ and the electric potential $A_{0}$ are the intensives variables ($I^{a}$). Hence, the coordinates of the thermodynamic phase space ${\cal T}$ in our case are parameterized by $Z^{A}=\{ M_{\rm eff}(S,\Lambda_{\rm eff}), S, \Lambda_{\rm eff}, T\}$, and the Gibbs 1-form in the mass representation can be expressed as
\begin{equation}
\Theta_{M}=dM_{\rm eff}-TdS-P d\Lambda_{\rm eff}\;.\label{fgs}
\end{equation}
So we have $\varphi^{*}(\Theta_{S})=0$, ensuring the existence of the first law of the thermodynamics of black holes, {\it viz.},  $dM_{\rm eff}=TdS+P d\Lambda_{\rm eff}$.

The non-degenerate metric (\ref{mt}) of the thermodynamic phase space ${\cal T}$ of the second order phase transition is given by 
\begin{eqnarray}
G & = &\left(dM_{\rm eff}-TdS-P d\Lambda_{\rm eff}\right)^{2}\nonumber\\&&+\left(TS+P \Lambda_{\rm eff}\right)\left[-dSdT+dP d\Lambda_{\rm eff}\right]\label{mt1}\;.
\end{eqnarray}
The pullback (\ref{me}) of the space of thermodynamic equilibrium states, taking into account (\ref{mass2}), takes the form 
\begin{widetext}
\begin{eqnarray}
g &=&\left(S\frac{\partial M_{\rm eff}}{\partial S}+\Lambda_{\rm eff}\frac{\partial M_{\rm eff}}{\partial \Lambda_{eff}}\right)\left(-\frac{\partial^2 M_{\rm eff}}{\partial S^2}dS^2+2\frac{\partial M_{\rm eff}}{\partial \Lambda_{\rm eff}}\frac{\partial M_{\rm eff}}{\partial S}d\Lambda_{\rm eff}dS\right)\;,\label{me1}\nonumber\\
&=&g_{SS}dS^2+2g_{\Lambda_{\rm eff}S}d\Lambda_{\rm eff}dS\,,\label{marng}
\end{eqnarray}
\end{widetext}
where the GTD metric coefficients reads
\begin{eqnarray}
g_{SS}&=&-\frac{5 \Lambda_{\rm eff}^2 S}{96 \pi ^3}
,\\
g_{\Lambda_{\rm eff}S}&=&\frac{5 \Lambda_{\rm eff} ^2 S^{7/2}}{144 \pi ^{9/2}}.
\end{eqnarray}
We will end our present analysis in this section with the study of the curvature singularities. To this end, we have to investigate scalars formed by the components of the curvature tensor. It is reasonable to construct the Ricci scalar $R=g^{ab}g^{cd}R_{acbd}$ which is the simplest scalar available. We finally discover that the scalar of curvature associated to the metric (\ref{marng}) takes the form
\begin{eqnarray}
&&R(S,\Lambda_{\rm eff})=\frac{432 \pi ^6S^6}{5 \Lambda_{\rm eff} ^4 }
\label{r1}
\end{eqnarray}
Finally, we obtain the following solutions for singularities for $R$:
\begin{eqnarray}
S=\infty. \label{get}
\end{eqnarray}
Surprisingly, we have found that this solution exactly matches that of the previous section on thermodynamics. We may stress here that a black hole with infinite entropy means no further change at that point can take place.

\section{Conclusion}
In this work, we revisited the modified theories of the scalar-vector-tensor gravity recently proposed by \cite{Heisenberg:2018acv}. Here by following an interesting guideline from \cite{Heisenberg:2018acv}, we have studied its cosmological and astronomical implications. We recapitulate our results as follows:
\begin{itemize}
\item New black hole solutions

We have firstly examined new black hole solutions which may be accommodated as one of the crucial implications of these SVT theories. Regarding this first possibility, we discovered the new metric function $h(r)$ written as
\begin{eqnarray}
h(r)= \frac{\Lambda _{\rm eff} r^2}{3}-\frac{2M_{\rm eff}}{r},\label{frc}
\end{eqnarray}
we have defined effective mass $M_{\rm eff}$ and cosmological constant $\Lambda_{\rm eff}$ in Sec.\ref{s3}. It is worth noting that the above metric function coincides with that of the planar de Sitter-Schwarzschild which corresponds to the gravitational field of a  non-rotating, spherically symmetric body of mass $M_{\rm eff}$ in de Sitter spacetime 

\item Thermodynamics

The thermodynamic implications are our next target. Here we followed the semi-classical approach to the thermodynamics of black holes. We computed the entropy of the new BH and found that entropy satisfied the usual area-law. we also quantified the Hawking temperature in terms of the obtained entropy. We observed that the temperature increases as a function of entropy $S$. Then, as the entropy increases, the temperature becomes an increasing function. We quantified the behavior of the temperature as a function of the entropy in Fig.\ref{BHtem}. 

Subsequently, we calculated the heat capacity and the possibility of a divergence point was pointed out. As is well-known, the divergence marks the phase transition point for the black holes. We have also checked thermal stability of the BH solutions by considering the sign of heat capacity, $C_{\Lambda_{\rm eff}}$. To this end, we wrote the heat capacity in terms of the entropy:
\begin{eqnarray}
C_{\Lambda_{\rm eff}}=2S, \label{BHcaps}
\end{eqnarray}
and observed the existence of the phase transition.
\item Geometrothermodynamics

Another interesting implication of the SVT theories in the present analysis arises when using the formalism of geometrothermodynamics. Using this approach, we have successfully constructed the metric of thermodynamic equilibrium states. This allowed us to explicitly determine the scalar curvature of the underlying theories. We also studied the curvature singularities and discovered that there exist thermodynamics interactions. In addition, the solution we obtained via this analysis is consistent with that of the usual thermodynamics. 

\end{itemize}

According to our result, we suggest that a further study of the SVT theories for other cosmological implications is reasonable. Likewise, implications for gravitational wave physics could provide some interesting phenomena. However, we will leave these interesting topics for our future projects.

\section*{Acknowledgements}
We thank Gregory W. Horndeski for his thorough readings and intuitive comments on our manuscript. We also thank Levinia Heisenberg for very useful comments and suggestions.


\end{document}